\documentclass[12pt]{article}
\usepackage[margin=25mm]{geometry}
\usepackage{amsmath}
\usepackage{graphicx,psfrag,epsf}
\usepackage{enumerate}
\usepackage{natbib}
\usepackage{amssymb}
\usepackage{epstopdf}
\usepackage{enumerate}
\usepackage{placeins}
\usepackage{setspace}
\usepackage{float}
\usepackage{soul,color}
\usepackage{placeins}
\usepackage{upgreek, mathtools,bm}
\usepackage{algorithm,algpseudocode,tabularx}
\usepackage{caption}
\usepackage{subcaption}
\usepackage{enumitem}
\usepackage{multirow}
\usepackage{amsfonts,amssymb,graphics,amsthm, comment}
\usepackage{url}
\RequirePackage[colorlinks,breaklinks,linkcolor=blue,citecolor=blue,urlcolor=blue]{hyperref}
\usepackage{multicol}
\usepackage{listings} 
\usepackage{xcolor} 
\immediate\write18{texcount -tex -sum  \jobname.tex > \jobname.wordcount.tex}

\graphicspath{{./plots/}}
\newtheorem{theorem}{Theorem}[section]

\newtheorem{proposition}[theorem]{Proposition}

\theoremstyle{remark}

\theoremstyle{plain}
\theoremstyle{definition}
\newtheorem{remark}{{\bf Remark}}
\newtheorem{assumption}{{\bf Assumption}}

\usepackage{booktabs}
\usepackage[flushleft]{threeparttable}
\usepackage{bbding,bbm,mathrsfs}

\newcommand\Prob{\mathbb{P}}

\DeclareMathOperator*{\argmin}{argmin} 
\DeclareMathOperator*{\argmax}{argmax}

\DeclareMathOperator{\E}{\textit{E}}
\def\T{^{ \mathrm{\scriptscriptstyle T} }}

\usepackage{xr-hyper}
\makeatletter

\newcommand*{\addFileDependency}[1]{
\typeout{(#1)}
\@addtofilelist{#1}
\IfFileExists{#1}{}{\typeout{No file #1.}}
}\makeatother

\providecommand{\keywords}[1]
{
  \small	
  \textbf{\textit{Keywords---}} #1
}

\title{Likelihood-based Nonparametric Receiver Operating Characteristic Curve Analysis in the Presence of Imperfect Reference Standard}
\author{Yifan Sun$^{1}$, Peijun Sang$^{1}$, Qinglong Tian$^{1}$, and Pengfei Li$^{1}$\\
        \small $^{1}$Department of Statistics and Actuarial Science, University of Waterloo \\
}
\date{} 

\begin{document}
\maketitle

\begin{abstract}
In diagnostic studies, researchers frequently encounter  imperfect reference standards with some misclassified labels. Treating these as gold standards can bias receiver operating characteristic (ROC) curve analysis.
To address this issue, we propose a novel likelihood-based method under a nonparametric density ratio model.
This approach enables the reliable estimation of the ROC curve, area under the curve (AUC), partial AUC, and Youden's index with favorable statistical properties. 
To implement the method, we develop an efficient expectation-maximization algorithm algorithm. Extensive simulations evaluate its finite-sample performance, showing smaller mean squared errors in estimating the ROC curve, partial AUC, and Youden's index compared to existing methods. We apply the proposed approach to a malaria study.
\end{abstract} \hspace{10pt}


\section{Introduction}\label{chap_intro}

The receiver operating characteristic (ROC) curve is a widely used tool for evaluating the performance of a continuous biomarker in distinguishing diseased individuals from healthy ones. 
Biomarker values are typically assumed to be higher in diseased individuals \citep{hu2023statistical}, with individuals classified as diseased if their biomarker value exceeds a threshold $t$. 
Under this assumption, sensitivity (i.e., true positive rate) is $1-F_1(t)$ and specificity (i.e., $1-$ false positive rate) is $F_0(t)$, where $F_0(t)$ and $F_1(t)$ are the cumulative distribution functions of the biomarker for healthy and diseased populations, respectively. 
The ROC curve plots sensitivity against $1-$ specificity across various thresholds. 
Mathematically, it is defined as 
\begin{equation}\label{ROC_def}
\textsc{roc}(s)=1- F_1\{ F_0^{-1}(1-s)\},\quad 
\mbox{for}\quad 0< s<1, 
\end{equation}
where $F_0^{-1}(1-s)=\inf\{t:F_0(t)\geq 1-s\}$. 

Several summary indices are commonly used with the ROC curve. The area under the ROC curve (AUC) and the partial AUC quantify the overall and partial discriminatory power, respectively. Youden's index \citep{youden1950index} represents the maximum effectiveness of a test compared to random guessing. 
For a comprehensive review of ROC analysis, see \citet{pepe2003statistical}, \citet{zhou2014statistical}, and references therein.

Estimation of ROC curves and related summary indices has been extensively studied, including parametric \citep{pepe2000interpretation}, nonparametric \citep{hsieh1996nonparametric, jokiel2013nonparametric}, and semiparametric methods \citep{qin2003using, zhou2008semi, yuan2021semiparametric}. 
Most methods rely on the availability of a gold standard for disease status ($G$: 1=diseased, 0=healthy). 
However, in many diagnostic studies, an individual’s disease status may not be definitively known. 
With the absence of a gold standard, an imperfect reference standard ($R$) is often used, where $R=1$ or $R=0$ indicates nominal classification as diseased or healthy, respectively. 
The reference standard $R$ is imperfect because it does not fully align with the unobserved underlying truth $G$. 
As a result, the nominal diseased group ($R = 1$) may include some truly healthy individuals, while the nominal healthy group ($R = 0$) may include some truly diseased individuals \citep{sun2024nonparametric}.
For further details on the imperfect reference standard, we refer readers to \cite{sun2024nonparametric} and Section \ref{chap_realdata}.

This paper focuses on nonparametric estimation of the ROC curve and related summary indices using two-sample data: one from the $R = 1$ group and the other from the $R = 0$ group, in which the gold standard $G$ is unavailable for both samples. 
In the literature, some efforts have been made to address situations where the gold standard for disease status is unavailable. For example, parametric Gaussian mixture models were explored by \citet{henkelman1990receiver} and \citet{hsieh2009interval}, while \citet{hall2003nonparametric} investigated nonparametric identifiability using multiple biomarkers.
Additionally, \citet{zhou2005nonparametric1} and \citet{albert2007random} examined ROC analysis for ordinal responses.
However, these approaches do not leverage the information provided by the reference standard $R$. They were developed for scenarios with no gold standard and do not address cases where an imperfect reference standard is available.

When a reference standard $R$ exists, treating $R$ as the gold standard $G$ and naively applying existing ROC analysis methods can reduce the power of a test, resulting in underestimation of the ROC curve and its related summary indices. 
In recent work,
within the two-sample framework mentioned above, one sample from the $R = 1$ group and the other from the $R = 0$ group, \citet{sun2024nonparametric} demonstrated that $F_0$, $F_1$, and $\textsc{roc}(s)$ in \eqref{ROC_def} are nonparametrically identifiable under the assumptions of conditional independence and the known accuracy of $R$ (see Assumption \ref{zhou-asmp} in Section \ref{chap_setting}). They developed a fully nonparametric approach for estimating the ROC curve and AUC while accounting for the reference standard, and analyzed the asymptotic properties of the proposed method. 

However, we note that the method and theoretical results developed in \cite{sun2024nonparametric} have two main limitations.  First, their approach for estimating $\textsc{roc}(s)$ relies on an inversion method to estimate  $F_0$ and $F_1$. This method does not take advantage of the continuity of the biomarker and lacks a likelihood-based interpretation, leaving potential for improvement. 
More details can be found in Section \ref{chap_setting}.
Second, \citet{sun2024nonparametric} did not consider the estimation of partial AUC and Youden’s index.

In this paper, we propose a likelihood-based nonparametric method to estimate $\textsc{roc}(s)$ and related summary indices within the two-sample framework mentioned above. 
To the best of our knowledge, this is the first likelihood-based nonparametric approach developed for analyzing continuous biomarkers in the presence of an imperfect reference standard in ROC analysis.
Our contributions are multifaceted. 
First, using the two-sample data, 
we construct a likelihood function with the log density ratio,
$\log\{f_1(t)/f_0(t)\}$, as the argument,
where $f_1(t)$ and $f_0(t)$ are the probability density functions corresponding to $F_1(t)$ and $F_0(t)$, respectively.  
This result extends the work of \citet{ward2009presence}, originally developed for presence-only data, a special case of the two-sample framework. 
Second, we propose a nonparametric density ratio model where $\log\{f_1(t)/f_0(t)\}$ is fully nonparametric and can be approximated by B-spline basis functions. The parameters in the B-spline approximation are estimated via the maximum likelihood method. 
This modeling strategy effectively incorporates information from both the $R=1$ and $R=0$ groups while leveraging the continuity of the biomarker. 
As demonstrated through simulations in Section \ref{chap_simu}, this approach may yield smoother and more efficient ROC curves compared to the method proposed by \citet{sun2024nonparametric}.
Third, using the estimated log density ratio, we propose estimators for ROC curve, AUC, partial AUC, and Youden’s index; we further establish the consistency of these estimators.
Fourth, we develop an efficient and stable expectation-maximization (EM) algorithm for numerical implementations. The convergence properties of the proposed algorithm are also established. 
The EM algorithm is particularly appealing because the E-step has a closed-form expression, and the M-step can be efficiently solved using existing well-tested R packages, avoiding direct optimization challenges.
Fifth, we conduct extensive simulations to compare the proposed methods with those in \cite{sun2024nonparametric}. The results demonstrate that the proposed methods outperform \cite{sun2024nonparametric}'s approach in estimating ROC curve, Youden's index, and partial AUC.

\section{Problem Setup and Review of the Literature}\label{chap_setting}
\subsection{Problem setup}
Recall that $G$ denotes the true disease status, which serves as the gold standard, with $G=1$ indicating a diseased subject and $G=0$ indicating a healthy subject. In this study, we assume that $G$ is unobserved, and only the imperfect reference standard $R$ is observable. We define the subjects with $R=1$ and $R=0$ as the nominal case group and the nominal control group, respectively.  

Define $$\pi_1=\Prob(G=1 \mid R=1),\quad \pi_0=\Prob(G=0 \mid R=0).$$  
The classical case-control study \citep{prentice1979logistic} corresponds to the special case where $\pi_1 = \pi_0 = 1$, meaning the gold standard is fully available. The study of presence-only data \citep{ward2009presence} or positive and unlabeled data \citep{song2019pulasso,liu2024positive} focuses on the case where $\pi_1 = 1$ and $\pi_0 \in (0,1]$, meaning the true status of the nominal case group can be perfectly determined, while that of the nominal control group remains undetermined. In this paper, we consider a more general and realistic setting where $\pi_0, \pi_1 \in (0,1]$. 

Let $T$ be a real-valued continuous biomarker with a bounded interval range $\mathcal{T}$. Since the value of biomarker is generally bounded in real world applications, without loss of generality, we assume $\mathcal{T} = [0,1]$. For any $t \in \mathcal{T}$, $F_0(t)$ and $F_1(t)$ are expressed as $F_0(t) = \Prob(T \le t \mid G = 0)$ and $F_1(t) = \Prob(T \le t \mid G = 1)$.

To ensure the identifiability of $F_0$ and $F_1$, we adopt the following assumption from \citet{sun2024nonparametric}:
\begin{assumption}\label{zhou-asmp}
(i) $\pi_1$ and $\pi_0$ are known, satisfying $\pi_0 + \pi_1 > 1$.
(ii) $R$ and $T$ are conditionally independent given $G$.
\end{assumption}
Part (i) of Assumption \ref{zhou-asmp} indicates that using the reference standard $R$ as a classifier to predict the gold standard $G$ is more informative than making a random guess. We assume that $\pi_1$ and $\pi_0$ are known; in practice, they may either be known or estimated from auxiliary data \citep{ma2012efficient,yu2019maximum}.
Part (ii) is a commonly adopted assumption in the literature \citep{hall2003nonparametric,zhou2005nonparametric1}. 
\citet{sun2024nonparametric} showed that $F_0$ and $F_1$ are unidentifiable if either part (i) or part (ii) of Assumption \ref{zhou-asmp} is violated.

Let $\{X_1, \dots, X_n\}$ and $\{Y_1, \dots, Y_m\}$ be two random samples of the biomarker $T$ from the nominal control group ($R=0$) and the nominal case group ($R=1$), respectively. Let $\{T_1, \dots, T_{n+m}\}$ denote the combined sample. Recall that $f_1(t)$ and $f_0(t)$ are the probability density functions corresponding to $F_1(t)$ and $F_0(t)$, respectively.
By Assumption \ref{zhou-asmp}(ii), we have:
\begin{equation}\label{mix_model}
X_1,\dots,X_n\sim f_0^*=\pi_0f_0+(1-\pi_0)f_1,\quad
Y_1,\dots,Y_m\sim f_1^*=(1-\pi_1)f_0+\pi_1f_1.
\end{equation}

The parameters of interest include the ROC curve, $\textsc{roc}(s)$, defined in \eqref{ROC_def}, and the related summary indices:  AUC,  partial AUC, and Youden's index. These are defined as follows:
\begin{equation}
\textsc{auc} = \int_{\mathcal{T}} \textsc{roc}(s)  \mathrm{d}s, \quad
\text{p}\textsc{auc} = (s_1 - s_0)^{-1} \int_{s_0}^{s_1} \textsc{roc}(s) \mathrm{d}s, \quad
J = \max_{t \in \mathcal{T}} \{F_0(t) - F_1(t)\}.
\label{summary_indices}
\end{equation}
Here, $s_0$ and $s_1$ in the partial AUC are prespecified values in the interval $[0,1]$.

When two or more biomarkers are available, a natural question is which biomarker can best differentiates between diseased and healthy individuals. To compare the performance of two biomarkers, $T_1$ and $T_2$, we assume that \eqref{mix_model} holds for each biomarker, and focus on the differences in $\textsc{roc}(s)$, $\textsc{auc}$, $\text{p}\textsc{auc}$, and $J$ between the two biomarkers.

\subsection{Overview of the methods in Sun et al. (2024)}\label{review_Sun}
In this section, we briefly review the fully nonparametric method
in \cite{sun2024nonparametric}. 
Note that \eqref{mix_model} implies that 
\begin{equation}\label{def_F01star_main}
    F_0^*(t)=\pi_0F_0(t)+(1-\pi_0)F_1(t),\quad
    F_1^*(t)=(1-\pi_1)F_0(t)+\pi_1F_1(t), \quad t\in\mathcal T,
\end{equation}
where $F_0^*$ and $F_1^*$ denote
the cumulative distribution functions of $T$ in 
$R=0$ group and $R=1$ group, respectively. 
By  solving $F_0$ and $F_1$ from the two equations in \eqref{def_F01star_main}, 
 we have
\begin{equation}\label{def_inversion}
    F_0(t)=\frac{\pi_1 F_0^*(t)-(1-\pi_0)F_1^*(t)}{\pi_0+\pi_1-1},\quad
    F_1(t)=\frac{\pi_0 F_1^*(t)-(1-\pi_1)F_0^*(t)}{\pi_0+\pi_1-1}, \quad t\in\mathcal T.
\end{equation}
Motivated by \eqref{def_inversion}, 
\citet{sun2024nonparametric}
proposed to estimate $F_0$ and $F_1$ by  
\begin{equation*}
    \tilde F_0(t)=\frac{\pi_1 \tilde F_0^*(t)-(1-\pi_0)\tilde F_1^*(t)}{\pi_0+\pi_1-1},\quad
    \tilde F_1(t)=\frac{\pi_0 \tilde F_1^*(t)-(1-\pi_1)\tilde F_0^*(t)}{\pi_0+\pi_1-1}, \quad t\in\mathcal T,
\end{equation*}
where $\tilde F_0^*$ and $\tilde F_1^*$ are the empirical distribution distributions based on the sample from $R=0$ and the sample from $R=1$, respectively. 
To  ensure that the estimated cumulative distribution functions are non-decreasing, a monotone transformation \citep{chernozhukov2010quantile} is further applied. 
We still denote the resulting estimators by $\tilde F_0(t)$ and $\tilde F_1(t)$, 
which can be plugged into \eqref{ROC_def} and \eqref{summary_indices}
to obtain the estimators of the ROC curve, AUC,  partial AUC, and Youden's index.
It is worth noting that \citet{sun2024nonparametric} only developed the asymptotic properties of the estimators of ROC curve and AUC, but did not consider those for the estimators of partial AUC and Youden's index.

Although easy to implement, \cite{sun2024nonparametric}'s method 
has two main limitations. 
First, it does not have a likelihood interpretation. 
Likelihood-based approaches are generally recognized for their efficiency compared to other methods. For parametric models, they are known to achieve efficiency under mild conditions. Moreover, they can achieve semiparametric efficiency in numerous semiparametric settings  
\citep{Bickel1993, kosorok2008introduction}.
Second, \cite{sun2024nonparametric}'s method does not take advantage of the continuity of the biomarkers and the smoothness of the density functions.
These limitations motivate us to develop a  likelihood-based nonparametric approach for estimating the ROC curve and associated summary indices, explicitly incorporating the continuous nature of  biomarkers.

\section{Method and Asymptotic Properties}\label{chap_loglik}

\subsection{Observed log-likelihood}\label{ward}
In this section, we adopt the case-control approach in \citet{McCullagh1989}
to develop the likelihood for the log density ratio: 
\begin{equation}\label{model_den.ratio}
h(t)=\log\left\{\frac{f_1(t)}{f_0(t)}\right\},\quad t\in\mathcal T.
\end{equation}
Specifically, we introduce an auxiliary binary random variable $S$
to indicate whether an observation is selected ($S=1$) or not ($S=0$) from the population. The introduction of $S$ is to account for different sampling rates in the nominal case and control groups. 
 We assume that $S$ only depends on $R$ \citep{McCullagh1989},  meaning that 
$S$ is independent of $G$ and $T$ conditionally on $R$. 
Given that the nominal control and case groups contain $m$ and $n$ observations, respectively,  the following relationship holds between the selection probabilities \citep{song2019pulasso}:
\begin{equation}
\frac{\Prob(R=1 \mid S=1)}
{\Prob(R=0 \mid S=1)}
=\frac{m}{n}.
\label{case_control_ratio}
\end{equation}
Consequently, the two-sample data from \eqref{mix_model} can be collectively represented as $\{T_i,R_i,S_i=1\}_{i=1}^{n+m}$.

By Bayes's formula and the conditional independence 
of $T$ and $S$ given $R$, 
we have 
\begin{eqnarray*}
\Prob(R=1\mid S=1,T=t)
&=&\frac{f_1^*(t)\Prob(R=1,S=1)}
{f_1^*(t)\Prob(R=1,S=1)+f_0^*(t)\Prob(R=0,S=1)}\\
&=&\frac{f_1^*(t)\Prob(R=1\mid S=1)}
{f_1^*(t)\Prob(R=1\mid S=1)+f_0^*(t)\Prob(R=0\mid S=1)},
\end{eqnarray*}
where $f_0^*$ and $f_1^*$ are defined in \eqref{mix_model}.
Using \eqref{mix_model} and \eqref{case_control_ratio}, we further obtain
\begin{eqnarray}
\label{posterior_prob}
\Prob(R=1\mid S=1,t)
\equiv
\Prob(R=1\mid S=1,T=t)
=\frac{(1-\pi_1)\lambda+\pi_1\lambda\exp \{h(t)\}}{1-\lambda_*+\lambda_*\exp \{h(t)\}},
\end{eqnarray}
where 
\begin{equation}\label{def:lambda_prop}
\lambda\equiv \frac m{n+m},\quad\lambda_*\equiv \frac{n(1-\pi_0)+m\pi_1}{n+m}.
\end{equation}
By \eqref{posterior_prob} and \eqref{def:lambda_prop}, 
the observed log-likelihood of $h$ 
based on $\{T_i,R_i,S_i=1\}_{i=1}^{n+m}$ is
\begin{equation}\label{lik_ward_2}
\begin{split}
l(h)&= \sum_{i=1}^{n+m}
\Big[R_i\log\{\Prob(R_i=1 \mid  S_i=1, T_i)\}+(1-R_i)\log\{\Prob(R_i=0 \mid   S_i=1, T_i)\}\Big]\\
&=\sum_{i=1}^{n+m}\Bigg(
R_i\log\left[\frac{(1-\pi_1)\lambda+\pi_1\lambda\exp \{h(T_i)\}}{1-\lambda_*+\lambda_*\exp \{h(T_i)\}}\right]\\
&\quad\qquad+(1-R_i)\log\left[\frac{\pi_0(1-\lambda)+(1-\pi_0)(1-\lambda)\exp \{h(T_i)\}}{1-\lambda_*+\lambda_*\exp \{h(T_i)\}}\right]\Bigg).
\end{split}\end{equation}

We make two comments regarding the observed log-likelihood function  
\eqref{lik_ward_2}. First, although we consider a one-dimensional biomarker  $T$, the result is also applicable for a multivariate $T$. 
Second, when $\pi_0=1$, 
\eqref{lik_ward_2} reduces to that of the presence-only model in \cite{ward2009presence}. Hence, our results generalize those in \cite{ward2009presence}.

\subsection{Point estimation\label{sec:ROC}}
We first discuss the estimation of the log density ratio function $h$. 
Since $h$ is infinite-dimensional, 
directly optimizing $l(h)$ in (\ref{lik_ward_2}) with respect to $h$
is infeasible.  
To address this, we employ the sieve estimation method.
In particular, 
let $K$ denote the number of B-spline basis functions with degree $d\ge1$. 
Let $0 = \tau_1<\cdots<\tau_{K-d+1} = 1$ be the knots of these B-spline basis functions. 
We approximate $h$ by the B-spline basis expansion: 
\[h(t)\approx   b\T \phi(t)=\sum_{k=1}^K b_k\phi_k(t), \quad t\in\mathcal T,\]
where $ \phi(t)=\{\phi_1(t),\dots,\phi_K(t)\}\T$ are the normalized B-spline basis functions over $\mathcal T$, and $  b=(b_1,\dots,b_K)\T$ are unknown coefficients to be estimated. 
Define 
\begin{equation}\label{obj_em_nopen}
\hat{  b}=\argmax_{  b\in \mathcal{B}_{K,B}}\, l(  b\T \phi), 
\end{equation}
where $l(\cdot)$ is defined by \eqref{lik_ward_2}, and $\mathcal{B}_{K,B}=\{  b\in\mathbb R^K:\|  b\|_\infty\le B\}$, with $\|b\|_\infty = \max_{1 \leq j \leq K}|b_j|$, and $B$ being a sufficiently large number.
We then obtain the estimator $\hat h (t)=\hat{  b}\T \phi(t)$. 
The numerical procedure to obtain $\hat h$ will be provided in Section \ref{chap_algo}. 

Establishing desirable  theoretical properties for $\hat{h}$  requires  a proper choice of $K$ and $B$, which may depend on $n$ and $m$, i.e., $K = K_{n, m}$ and $B = B_{n, m}$.
More details can be found in Condition 2 
in the supplementary material. For simplicity, 
we omit the subscripts $n,~m$ in the  notation. 

With the estimated log density ratio $\hat h$, we now discuss the estimation of $F_0(t)$ and $F_1(t)$.
Recall $\lambda_*$ in \eqref{def:lambda_prop}, 
which represents the proportion of true diseased individuals in the combined sample  $\{T_1,\ldots,T_{n+m}\}$.
Let $H(t)=(1-\lambda_*) F_0(t)+\lambda_* F_1(t)$,
which can be viewed as the cumulative distribution function of  $\{T_1,\ldots,T_{n+m}\}$. 
We use $\E_H$ to denote the expectation with respect to $H(t)$. 
Using \eqref{model_den.ratio}, it is straightforward to verify that 
\begin{equation}\label{prop_P_R_F_eq}
\E_H\left\{\frac{\mathbbm 1(T\le t)}{(1-\lambda_*) +\lambda_* \exp \{h(T)\}}\right\}=F_0(t),
\; 
\E_H\left\{\frac{\mathbb \exp \{h(T)\}\mathbbm 1(T\le t)}{(1-\lambda_*) +\lambda_* \exp \{h(T)\}}\right\}=F_1(t),\; t\in\mathcal T.
\end{equation}
Here $\mathbbm 1(\cdot)$ is the indicator function. 

Equation \eqref{prop_P_R_F_eq} provides insight into the estimation of $F_0(\cdot)$ and  $F_1(\cdot)$. 
For $i=1,\dots,n+m$, let 
$$
\hat p_i= (n+m)^{-1} [1-\lambda_*+\lambda_*\exp \{\hat h(T_i)\}]^{-1}.
$$
Motivated by \eqref{prop_P_R_F_eq}, 
we propose the following H\'ajek-type estimators \citep{wu2020sampling} for $F_0$ and $F_1$:    
\begin{equation}\label{Fest.practice0}
\hat F_0(t)=\frac{\sum_{i=1}^{n+m}\hat p_i \mathbbm 1(T_i\le t)}{\sum_{i=1}^{n+m} \hat p_i}, 
\quad
\hat F_1(t)=\frac{\sum_{i=1}^{n+m}\hat p_i \exp \{\hat h(T_i)\} \mathbbm 1(T_i\le t)}{\sum_{i=1}^{n+m} \hat p_i \exp \{\hat h(T_i)\}}, \quad t\in\mathcal T. 
\end{equation}

Using \eqref{Fest.practice0}, 
we estimate the ROC curve, the AUC, and the partial AUC as follows:    
\begin{eqnarray}\label{ROC_est}
\widehat{\textsc{roc}}(s)&=&1-\hat F_1\{\hat F_0^{-1}(1-s)\},\,\,\,
\widehat{\textsc{auc}}=\int_{0}^{1} \widehat{\textsc{roc}}(s)\mathrm ds, \,\,\,
\widehat{\text{p}\textsc{auc}}=\frac{1}{s_1-s_0}\int_{s_0}^{s_1} \widehat{\textsc{roc}}(s)\mathrm ds.
\end{eqnarray}
To estimate  Youden's index $J$, define the optimal cutoff $C_0=\argmin_{t\in\mathcal T}\{F_0(t)-F_1(t)\}$. Clearly, $C_0$ satisfies  $f_0(C_0)=f_1(C_0)$, which is equivalent to $h(C_0)=0$ by the definition of $h$ in (\ref{model_den.ratio}). 
Hence, given $\hat h$, we can estimate $C_0$ by solving $\hat h(\hat C)=0$. 
If multiple solutions exist, we select the one that maximizes $\hat F_0(\cdot)-\hat F_1(\cdot)$.  
 Youden's index is then estimated by 
\begin{equation}\label{YI_est}
\hat J=\hat F_0(\hat C)-\hat F_1(\hat C).
\end{equation}


For the comparison between two biomarkers $T_1$ and $T_2$, we apply the proposed procedure to each biomarker and obtain $\widehat{\textsc{roc}}_1(s)$ and $\widehat{\textsc{roc}}_2(s)$. The difference $\Delta\textsc{roc}(s)
={\textsc{roc}}_1(s)-{\textsc{roc}}_2(s)$ is then estimated by $\widehat{\textsc{roc}}_1(s)-\widehat{\textsc{roc}}_2(s)$. We similarly estimate $\Delta\textsc{auc}={\textsc{AUC}}_1(s)-{\textsc{AUC}}_2(s)$, $\Delta\text{p}\textsc{auc}
=\text{p}\textsc{auc}_1-\text{p}\textsc{auc}_2$, and $\Delta J=J_1-J_2$.

We conclude this section with some comments on the proposed estimation methods.
First, we approximate the log density ratio $h(t)$ using B-spline basis functions; this approximation takes into account the continuity of the biomarkers and the smoothness of the density functions. In contrast, \citet{sun2024nonparametric} did not integrate these aspects into their procedure.
We also develop a likelihood-based method to combine the information from both the 
 $R=1$ and $R=0$ groups. 
We expect that the resulting ROC curve will be smoother and more efficient, as demonstrated in Section \ref{chap_simu}.
Second, the proposed modeling strategy is closely related to the density ratio model, or the exponential tilting model  \citep{qin1998inferences,qin2017biased},
in which $h(t)$ is modeled by a prespecified parametric form.  However, we model  $h(t)$ nonparametrically, thus avoiding the potential issue of model misspecification.


\subsection{Asymptotic properties}\label{chap_thm}
In this section, we establish the consistency of the proposed estimators in \eqref{Fest.practice0}--\eqref{YI_est}. 
For ease of presentation, the regularity conditions and proofs are provided in Sections S4--
S6 
  of the supplementary material. 
\begin{theorem}\label{thm:Fconsis}
Assume Assumption \ref{zhou-asmp} and 
model \eqref{mix_model} hold. 
\begin{enumerate}[label=(\roman*)]
\item 
If Conditions 1--3 
in the supplementary material are satisfied,   we have 
\[\sup_{t\in\mathcal T}|\hat F_0(t)-F_0(t)|=o_p(1),
\quad
\sup_{t\in\mathcal T}|\hat F_1(t)-F_1(t)|=o_p(1).\]
\item If, in addition,  Condition 4 
holds,
\begin{align*}\sup_{s\in[0,1]}|\widehat{\textsc{roc}}(s)-\textsc{roc}(s)|=o_p(1), 
~|\widehat{\textsc{auc}}-{\textsc{auc}}|=o_p(1),~|\widehat{\text{p}\textsc{auc}}-\text{p}\textsc{auc}|=o_p(1).
\end{align*}
\item If Conditions 1--6  
are all satisfied, then $|\hat J-J|=o_p(1)$. 
\end{enumerate}
\end{theorem} 

\section{Computing Algorithm}\label{chap_algo}

\subsection{Penalized likelihood}\label{sec:penLL}

To perform ROC curve analysis, solving the optimization problem \eqref{obj_em_nopen} is crucial. 
Despite the finite-dimensional nature, this problem remains challenging. 
First, the number of basis $K$ is often chosen to be sufficiently large to ensure that the log density ratio $h$ can be well approximated by the B-spline basis functions. 
However, without proper regularization, the resulting estimated $h$ may exhibit excessive variability when $K$ is large. 
Secondly, the objective function 
 $l(b\T\phi)$ in (\ref{obj_em_nopen}) is not a concave function in the B-spline coefficients $b$, which presents significant challenges in the maximization process. 

To address the first issue, 
we employ the penalized spline approach proposed by \cite{green1993nonparametric}. In particular, we consider a penalized version of \eqref{obj_em_nopen}:
\begin{equation}\label{obj_em_pen}
\max_{  b \in\mathbb R^K} Q(  b)=\max_{  b \in\mathbb R^K} \left\{l(  b\T \phi)-P_\nu(  b)\right\}, 
\end{equation}
where $\nu > 0$ is a tuning parameter, and 
$P_\nu(  b)=\nu\int_{\mathcal T} [\{  b\T \phi(t)\}^{(2)}]^2\mathrm dt$ is the penalty term, 
with $f^{(2)}$ representing the second derivative of the function $f$. 
This roughness penalty term 
can be rewritten as $\nu  b\T  \Phi_2   b$, 
where $\Phi_2 \in \mathbb{R}^{K \times K}$ with the $(j,k)$ element given by $\int_{\mathcal T}\phi^{(2)}_j(t)\phi^{(2)}_k(t)\mathrm dt$.

The tuning parameter 
$\nu$ can be selected through cross-validation, with further details provided in Section S8.2  
 of the supplementary material.

\subsection{EM algorithm}\label{sec:EMalgo}

The penalized log-likelihood in (\ref{obj_em_pen}) remains non-concave with respect to $b$, making direct maximization of \eqref{obj_em_pen} potentially unstable. To address this challenge,
we treat the unknown labels $\{G_i:i=1,\dots,n+m\}$ as missing data and adopt the EM algorithm framework \citep{dempster1977maximum} to develop more robust algorithms. 
 We first derive the penalized complete log-likelihood function under the assumption that
  $G_i$'s are known. 
For ease of presentation, the proofs in this section are provided in 
Sections S7 and S9 
of the supplementary material.
  
\begin{proposition}\label{prop_full-loglik}
Under Assumption \ref{zhou-asmp} and model (\ref{mix_model}), the complete log-likelihood function of $h$, based on $\{T_i,R_i,G_i,S_i=1\}_{i=1}^{n+m}$, is given by
\begin{equation*}
l^f(h)
=\sum_{i=1}^{n+m} \left(G_i\log\left[ \frac{\exp \{\tilde h(T_i)\}}{\exp \{\tilde h(T_i)\}+1} \right]
+(1-G_i)\log\left[ \frac{1}{\exp \{\tilde h(T_i)\}+1} \right]\right),
\end{equation*}
where $\tilde h(t)\equiv h(t)+c$ for $t\in\mathcal T$, with $c\equiv\log\{\lambda_*/(1-\lambda_*)\}$. 
Hence, the penalized complete log-likelihood function with respect to $b$ is 
$$
Q^f( b)=l^f(  b\T \phi)-P_\nu(  b).
$$ 
\end{proposition}


The EM algorithm consists of the E-step and the M-step. 
The E-step computes the expectation of $Q^f( b)$ conditional on the observed data and the current estimate of $h$. 
Equivalently, we impute $G_i$ in $Q^f( b)$ with the expectation of $G_i$ conditional on the observed data $\mathbb O=\{(R_i,T_i,S_i=1):i=1,\dots, n+m\}$, and the current estimated $h$. 
We next present the closed form of the E-step. 
Denote by $1_K$ the vector of $K$ ones. 

\begin{proposition}\label{prop_Estep}
Let $\tilde b= b- c1_K$ for any $ b\in\mathbb R^K$ with $c$ given in Proposition \ref{prop_full-loglik}, and denote by $h^{[s]}$ the estimated $h$ at the $s$th iteration step. Re-parameterizing $\E\{Q^f( b)\mid\mathbb O;h^{[s]}\}$ with respect to $\tilde b$, we have 
\begin{eqnarray*}
\tilde Q^{[s]}(\tilde b)
&=&\E\{Q^f( b)\mid\mathbb O;h^{[s]}\}\\
&=&\sum_{i=1}^{n+m} \left(\hat G^{[s]}_i\log\left[ \frac{\exp \{\tilde b\T \phi(T_i)\}}{\exp \{\tilde b\T \phi(T_i)\}+1} \right]
+(1-\hat G^{[s]}_i)\log\left[ \frac{1}{\exp \{\tilde b\T \phi(T_i)\}+1} \right]\right)-P_\nu(\tilde{  b}),
\end{eqnarray*}
where, for $i=1,\dots, n+m$,  
\begin{equation}\label{imputation0}
\hat G^{[s]}_i=\left[\frac{(1-\pi_0)\exp\{h^{[s]}(T_i)\}}{(1-\pi_0)\exp\{h^{[s]}(T_i)\}+\pi_0}\right]^{1-R_i}
\left[\frac{\pi_1\exp\{h^{[s]}(T_i)\}}{\pi_1\exp\{h^{[s]}(T_i)\}+1-\pi_1}\right]^{R_i}.
\end{equation}
\end{proposition} 

Using Proposition \ref{prop_Estep},
we present
the pseudo-code of the EM algorithm in Algorithm \ref{algo-EM-1}.  
\begin{algorithm}
\caption{EM algorithm for \eqref{obj_em_pen}.}\label{algo-EM-1}
\begin{algorithmic}[1]
\State Set $\nu>0$, and calculate $c$ as in Proposition \ref{prop_full-loglik}
\State Initialize $\hat G_i^{[0]}=(1-\pi_0)(1-R_i)+\pi_1R_i$ for $i=1,\ldots,n+m$ 
\State Define $h^{[1]}(t)=\phi\T(t)\tilde{  b}^{[1]}-c$ with $\tilde{  b}^{[1]}=\argmax_{\tilde b\in\mathbb R^K} \tilde Q^{[0]}(\tilde b)$
 \For {$s= 1,2,\ldots,$ }
\State E-step: \quad\, update $\hat G_i^{[s]}$ by \eqref{imputation0} and obtain $\tilde Q^{[s+1]}(\tilde b)$ 
\State M-step: \quad $\tilde b^{[s+1]}\leftarrow \argmax_{\tilde b\in\mathbb R^K} \tilde Q^{[s]}(\tilde b)$
\State Shifting: \,\,\;$ b^{[s+1]} \leftarrow \tilde b^{[s+1]}-c$;\quad $h^{[s+1]}(t)\leftarrow\phi\T(t)\tilde{  b}^{[s+1]}-c$
 \EndFor
\State  Output $h^{[s_0]}$ for some $s_0$ when the stopping criterion is met
\end{algorithmic}
\end{algorithm}

The stopping criteria are detailed in Section S8.1  
of the supplementary material. The initial value of $\hat{G}_i^{[0]}$ is selected based on the expectations of $G$ given $R=0$ and $R=1$, which are $1-\pi_0$ and $\pi_1$, respectively, as described in model (\ref{mix_model}).

\begin{remark}\label{rmk-mgcv}
The objective function $\tilde Q^{[s]}(\tilde b)$ in Proposition \ref{prop_Estep} is concave with respect to $\tilde{b}$ because the matrix $\Phi_2$ in $P_\nu$ is positive semi-definite. As a result, the M-step is straightforward to implement. Furthermore, $\tilde Q^{[s]}(\tilde b)$ resembles a nonparametric logistic regression with a roughness penalty, allowing us to leverage the R function \texttt{gam} from the \texttt{mgcv} package \citep{wood2017generalized} to simplify the optimization process, with $\hat{G}_i^{[s]}$ treated as the pseudo-binomial response.
\end{remark}

\begin{proposition}\label{em_thm}
For the sequence ${b^{[s]} : s=1,\dots}$ generated by Algorithm \ref{algo-EM-1}, we have
$Q(b^{[s+1]}) \geq Q(b^{[s]})$ for all $s \geq 0$, where $Q$ is defined in \eqref{obj_em_pen}.
Moreover, the inequality is strict if $\nabla Q(b^{[s]}) \neq 0$, where $\nabla Q(b)$ denotes the gradient of $Q(b)$ with respect to $b$.
\end{proposition} 

Proposition \ref{em_thm} provides a theoretical guarantee for the proposed EM algorithm. Specifically, it follows from (\ref{lik_ward_2}) that $l(h) \leq 0$ for any $h$. As a result, $Q(b)$ is bounded above by 0 for any $b$, since $\Phi_2$ in (\ref{obj_em_pen}) is positive semi-definite. Therefore, the ascending property established in Proposition \ref{em_thm} ensures that the values of the objective function $Q(b^{[s]})$ converge to a (local) maximizer of $Q(b)$ as $s \to \infty$.


\section{Simulation Study}\label{chap_simu} 
\subsection{General setup}

We conduct simulation studies involving both univariate and bivariate biomarkers, with settings similar to those in \citet{sun2024nonparametric}.
Let $\pi=\Prob(G=1)=0.4$.
The sensitivity and specificity are defined as \( \text{se} = \Prob(R = 1 \mid G = 1) \) and \( \text{sp} = \Prob(R = 0 \mid G = 0) \), respectively. 
We consider sample sizes of $n = m = 100$, 300, and 500, with values of se and sp set to 0.95, 0.90, or 0.75.
The corresponding values of $\pi_0$ and $\pi_1$, calculated using Bayes' formula, are displayed in Table S.1  
in the supplementary material. Each experiment is repeated 1000 times.
In the main text, we present results for normally distributed biomarkers, while results for gamma-distributed biomarkers are deferred to Section S.2  
in the supplementary material.

The number of B-spline basis functions (for each biomarker), denoted as $K$, is set to 50, and the tuning parameter $\nu$ is selected via five-fold cross-validation.
An additional study demonstrates that the estimates of the target parameters are insensitive to the number of B-spline basis functions.
We choose $d$, the degree of the B-spline basis functions, to be 4.

The parameters of interest include the ROC curve evaluated at 0.2, the AUC, the partial AUC from $s_0 = 0.1$ to $s_1 = 0.3$, and the Youden’s index. A similar partial AUC was considered by \citet{wan2008comparing}.
For bivariate biomarkers, we focus on the differences in these parameters, as outlined in Section \ref{sec:ROC}.
For convenience, Table S.2 
 in the supplementary material summarizes the true values of these parameters of interest.

\subsection{Univariate case}\label{simu:uni}  

Conditioned on $G=0$ or $G=1$, the biomarkers are generated from Gaussian distributions, $\mathcal{N}(0,1)$ and $\mathcal{N}(1,1)$, respectively. The corresponding log-density ratio is given by $h(t) = t - 1/2$. Two-sample data are then generated according to model (\ref{mix_model}) with mixture proportions $\pi_0$ and $\pi_1$.

We compare the proposed method with the fully nonparametric method \citep{sun2024nonparametric}, reviewed in Section \ref{review_Sun}, and with a naive method that treats $R$ as the gold standard $G$. Specifically, the naive method applies the proposed approach with $\pi_0 = \pi_1 = 1$, which are misspecified values.
The performance of each method is assessed based on the estimated bias, standard error, and the mean squared error of the estimates across 1000 Monte Carlo runs.

Table \ref{tab:uni-normal-1} presents the simulated results for the three methods under various combinations of sample sizes and misclassification levels. All methods show deteriorating performance as sensitivity and specificity decrease, while they perform better as sample sizes increase.
The proposed method significantly outperforms the fully nonparametric method in terms of mean squared errors for both the estimated ROC curve and the estimated Youden's index across all simulation settings. Notably, it consistently exhibits smaller standard errors than the fully nonparametric method. 
Additionally, the bias in the estimated Youden’s index using the fully nonparametric method is substantially larger than that of our method.
For estimating the AUC, the two methods show similar performance. 
However, the mean squared error of the estimated partial AUC using our method is lower than that of the fully nonparametric method.
Although the naive method has the smallest standard error among the three methods, neglecting the mixture structure in the naive method leads to a significantly larger bias and mean squared error compared to the other two methods across all simulation settings.

\begin{table}
	\caption{Simulated results for univariate normal settings in Section \ref{simu:uni}}
\small{	\setlength\tabcolsep{4pt}
    \renewcommand\arraystretch{0.8}
    \begin{tabular}{ccc|ccc|ccc|ccc|ccc}
    \toprule
          &       &       & \multicolumn{3}{c|}{ROC(0.2)} & \multicolumn{3}{c|}{AUC} & \multicolumn{3}{c|}{Youden} & \multicolumn{3}{c}{pAUC} \\
   se$\,=\,$sp    & $n=m$     & method & bias  & sd    & mse   & bias  & sd    & mse   & bias  & sd    & mse   & bias  & sd    & mse \\
    \midrule
    0.95  & 100   & EM    & -0.41  & 6.87  & 0.47  & -0.18  & 3.78  & 0.14  & 0.19  & 6.19  & 0.38  & -0.40  & 6.72  & 0.45  \\
          &       & NP    & 0.17  & 8.17  & 0.67  & -0.03  & 3.83  & 0.15  & 4.60  & 6.51  & 0.64  & 0.12  & 7.34  & 0.54  \\
          &       & naive & -5.02  & 6.01  & 0.61  & -2.88  & 3.40  & 0.20  & -4.05  & 5.47  & 0.46  & -4.91  & 5.85  & 0.58  \\
\cmidrule{2-15}          & 300   & EM    & -0.15  & 4.21  & 0.18  & -0.09  & 2.31  & 0.05  & 0.03  & 3.77  & 0.14  & -0.15  & 4.11  & 0.17  \\
          &       & NP    & 0.00  & 4.92  & 0.24  & -0.03  & 2.32  & 0.05  & 2.37  & 3.95  & 0.21  & 0.05  & 4.45  & 0.20  \\
          &       & naive & -4.94  & 3.66  & 0.38  & -2.90  & 2.06  & 0.13  & -4.35  & 3.31  & 0.30  & -4.84  & 3.56  & 0.36  \\
\cmidrule{2-15}          & 500   & EM    & 0.06  & 3.17  & 0.10  & 0.04  & 1.72  & 0.03  & 0.17  & 2.81  & 0.08  & 0.06  & 3.10  & 0.10  \\
          &       & NP    & 0.24  & 3.76  & 0.14  & 0.07  & 1.73  & 0.03  & 1.97  & 3.13  & 0.14  & 0.23  & 3.44  & 0.12  \\
          &       & naive & -4.81  & 2.76  & 0.31  & -2.83  & 1.53  & 0.10  & -4.30  & 2.46  & 0.25  & -4.72  & 2.68  & 0.29  \\
    \midrule
    0.9   & 100   & EM    & -0.46  & 8.02  & 0.65  & -0.31  & 4.45  & 0.20  & 0.17  & 7.24  & 0.52  & -0.45  & 7.84  & 0.62  \\
          &       & NP    & 0.32  & 8.99  & 0.81  & -0.03  & 4.55  & 0.21  & 5.18  & 7.52  & 0.83  & 0.29  & 8.32  & 0.69  \\
          &       & naive & -9.44  & 6.05  & 1.26  & -5.65  & 3.56  & 0.45  & -8.16  & 5.62  & 0.98  & -9.25  & 5.86  & 1.20  \\
\cmidrule{2-15}          & 300   & EM    & -0.15  & 4.96  & 0.25  & -0.11  & 2.72  & 0.07  & 0.07  & 4.38  & 0.19  & -0.14  & 4.84  & 0.23  \\
          &       & NP    & 0.04  & 5.62  & 0.32  & 0.00  & 2.74  & 0.08  & 2.75  & 4.54  & 0.28  & 0.09  & 5.15  & 0.27  \\
          &       & naive & -9.35  & 3.70  & 1.01  & -5.64  & 2.14  & 0.36  & -8.46  & 3.34  & 0.83  & -9.16  & 3.59  & 0.97  \\
\cmidrule{2-15}          & 500   & EM    & 0.05  & 3.78  & 0.14  & 0.03  & 2.04  & 0.04  & 0.21  & 3.33  & 0.11  & 0.05  & 3.69  & 0.14  \\
          &       & NP    & 0.29  & 4.39  & 0.19  & 0.08  & 2.06  & 0.04  & 2.32  & 3.60  & 0.18  & 0.30  & 4.00  & 0.16  \\
          &       & naive & -9.31  & 2.83  & 0.95  & -5.61  & 1.61  & 0.34  & -8.47  & 2.57  & 0.78  & -9.13  & 2.74  & 0.91  \\
    \midrule
    0.75  & 100   & EM    & -0.90  & 13.25  & 1.76  & -0.94  & 7.56  & 0.58  & -0.02  & 12.15  & 1.48  & -0.88  & 13.00  & 1.70  \\
          &       & NP    & 0.12  & 14.88  & 2.21  & 0.01  & 7.99  & 0.64  & 7.93  & 11.88  & 2.04  & 0.28  & 13.75  & 1.89  \\
          &       & naive & -21.10  & 5.75  & 4.78  & -13.51  & 3.83  & 1.97  & -19.46  & 5.81  & 4.13  & -20.64  & 5.53  & 4.57  \\
\cmidrule{2-15}          & 300   & EM    & -0.44  & 8.30  & 0.69  & -0.46  & 4.66  & 0.22  & -0.08  & 7.44  & 0.55  & -0.41  & 8.12  & 0.66  \\
          &       & NP    & 0.07  & 9.18  & 0.84  & 0.02  & 4.80  & 0.23  & 4.67  & 7.56  & 0.79  & 0.14  & 8.51  & 0.72  \\
          &       & naive & -21.04  & 3.56  & 4.55  & -13.53  & 2.29  & 1.88  & -19.93  & 3.52  & 4.10  & -20.56  & 3.42  & 4.34  \\
\cmidrule{2-15}          & 500   & EM    & -0.11  & 6.22  & 0.39  & -0.17  & 3.47  & 0.12  & 0.15  & 5.53  & 0.31  & -0.11  & 6.08  & 0.37  \\
          &       & NP    & 0.29  & 7.02  & 0.49  & 0.09  & 3.53  & 0.12  & 3.80  & 5.73  & 0.47  & 0.31  & 6.45  & 0.42  \\
          &       & naive & -21.09  & 2.60  & 4.52  & -13.52  & 1.70  & 1.86  & -20.02  & 2.63  & 4.08  & -20.62  & 2.50  & 4.31  \\
    \bottomrule
    \end{tabular}
}
	\label{tab:uni-normal-1}
	\begin{tablenotes}
    \item Se, sp, sensitivity and specificity; $n,m$, sample sizes; 
    EM, the proposed method; NP, the fully nonparametric method \citep{sun2024nonparametric}; 
    bias, sd, mse: {the estimated bias, standard error, and mean squared error across 1000 Monte Carlo simulations}; 
    ROC(0.2), the receiver operating characteristic curve value at 0.2; 
    AUC, the area under curve; 
    Youden, Youden's index; 
    pAUC, the partial AUC integrated over $[0.1,0.3]$. 
    All values have been multiplied by $10^2$.
	\end{tablenotes}
	\label{tab-note}
\end{table}

\subsection{Bivariate case}\label{simu:bi}  

Conditional on $G=0$ and $G=1$, the distributions of  bivariate biomarkers are, respectively, 
\begin{equation*}
\mathcal N\left(\left[
\begin{array}{c}
2\\1
\end{array}
\right],
\left[
\begin{array}{cc}
1 & \rho \\
\rho & 1
\end{array}
\right]
\right), \quad 
\mathcal N\left(\left[
\begin{array}{c}
0\\0
\end{array}
\right],
\left[
\begin{array}{cc}
1 & \rho \\
\rho & 1
\end{array}
\right]
\right),
\end{equation*}
where the correlation between the two biomarkers, denoted by
$\rho$, is 0.2. 
All other simulation settings remain identical to those described in Section \ref{simu:uni}.

Table \ref{tab:bi-normal-1} summarizes the differences in the estimated parameters across all simulation settings for the three methods. Similar patterns observed in Table \ref{tab:uni-normal-1} are also evident in Table \ref{tab:bi-normal-1}. For the mean squared errors of the estimated AUC, the proposed method shows a slight improvement over the fully nonparametric method when sample sizes are small.
However, the advantage of the proposed method is less pronounced than in the univariate settings, particularly in terms of the bias in the difference of the estimated Youden’s indices. One possible explanation is that the biases of the estimated Youden’s indices for these two biomarkers are of the same direction in most simulation settings, which results in their counteraction after subtraction.
This phenomenon is further explored in Section S3 
 of the supplementary material, where we present the marginal results for the two biomarkers separately.

\begin{table}
	\caption{Simulated results for bivariate normal settings in Section \ref{simu:bi}}
\small{	\setlength\tabcolsep{4pt}
    \renewcommand\arraystretch{0.8}
    \begin{tabular}{ccc|ccc|ccc|ccc|ccc}
    \toprule
          &       &       & \multicolumn{3}{c|}{$\Delta$ROC(0.2)} & \multicolumn{3}{c|}{$\Delta$AUC} & \multicolumn{3}{c|}{$\Delta$Youden} & \multicolumn{3}{c}{$\Delta$pAUC} \\
    se$\,=\,$sp    & $n=m$      & method & bias  & sd    & mse   & bias  & sd    & mse   & bias  & sd    & mse   & bias  & sd    & mse \\
    \midrule
    0.95  & 100   & EM    & -0.53  & 7.81  & 0.61  & -0.31  & 4.08  & 0.17  & -0.48  & 7.42  & 0.55  & -0.56  & 7.68  & 0.59  \\
          &       & NP    & -0.23  & 9.18  & 0.84  & -0.06  & 4.16  & 0.17  & -1.70  & 7.61  & 0.61  & -0.37  & 8.30  & 0.69  \\
          &       & naive & -3.78  & 7.02  & 0.64  & -1.89  & 3.67  & 0.17  & -4.75  & 6.48  & 0.65  & -4.06  & 6.88  & 0.64  \\
\cmidrule{2-15}          & 300   & EM    & -0.38  & 4.47  & 0.20  & -0.22  & 2.38  & 0.06  & -0.29  & 4.34  & 0.19  & -0.39  & 4.40  & 0.20  \\
          &       & NP    & -0.32  & 5.32  & 0.28  & -0.10  & 2.43  & 0.06  & -0.88  & 4.68  & 0.23  & -0.29  & 4.79  & 0.23  \\
          &       & naive & -3.77  & 4.19  & 0.32  & -1.92  & 2.17  & 0.08  & -4.66  & 4.01  & 0.38  & -4.01  & 4.12  & 0.33  \\
\cmidrule{2-15}          & 500   & EM    & -0.19  & 3.63  & 0.13  & -0.08  & 1.88  & 0.04  & -0.17  & 3.47  & 0.12  & -0.19  & 3.56  & 0.13  \\
          &       & NP    & -0.14  & 4.30  & 0.18  & -0.01  & 1.88  & 0.04  & -0.64  & 3.78  & 0.15  & -0.19  & 3.85  & 0.15  \\
          &       & naive & -3.73  & 3.38  & 0.25  & -1.86  & 1.68  & 0.06  & -4.69  & 3.17  & 0.32  & -3.97  & 3.30  & 0.27  \\
    \midrule
    0.9   & 100   & EM    & -0.86  & 8.77  & 0.78  & -0.41  & 4.70  & 0.22  & -0.59  & 8.46  & 0.72  & -0.87  & 8.65  & 0.76  \\
          &       & NP    & -0.27  & 10.04  & 1.01  & 0.08  & 4.81  & 0.23  & -1.78  & 8.43  & 0.74  & -0.35  & 9.12  & 0.83  \\
          &       & naive & -7.26  & 7.04  & 1.02  & -3.51  & 3.78  & 0.27  & -8.21  & 6.45  & 1.09  & -7.68  & 6.86  & 1.06  \\
\cmidrule{2-15}          & 300   & EM    & -0.46  & 5.20  & 0.27  & -0.28  & 2.76  & 0.08  & -0.39  & 5.10  & 0.26  & -0.47  & 5.12  & 0.26  \\
          &       & NP    & -0.31  & 6.08  & 0.37  & -0.11  & 2.84  & 0.08  & -1.08  & 5.37  & 0.30  & -0.40  & 5.46  & 0.30  \\
          &       & naive & -7.28  & 4.58  & 0.74  & -3.65  & 2.24  & 0.18  & -8.29  & 4.28  & 0.87  & -7.70  & 4.47  & 0.79  \\
\cmidrule{2-15}          & 500   & EM    & -0.26  & 4.14  & 0.17  & -0.14  & 2.17  & 0.05  & -0.22  & 3.99  & 0.16  & -0.26  & 4.06  & 0.17  \\
          &       & NP    & -0.12  & 4.77  & 0.23  & 0.01  & 2.15  & 0.05  & -0.75  & 4.18  & 0.18  & -0.19  & 4.32  & 0.19  \\
          &       & naive & -7.24  & 3.67  & 0.66  & -3.57  & 1.72  & 0.16  & -8.25  & 3.35  & 0.79  & -7.65  & 3.56  & 0.71  \\
    \midrule
    0.75  & 100   & EM    & -2.82  & 14.65  & 2.23  & -1.39  & 8.08  & 0.67  & -2.19  & 14.37  & 2.11  & -3.01  & 14.54  & 2.21  \\
          &       & NP    & -1.43  & 16.12  & 2.62  & 0.17  & 8.71  & 0.76  & -3.22  & 13.55  & 1.94  & -1.80  & 14.76  & 2.21  \\
          &       & naive & -17.88  & 7.59  & 3.77  & -8.45  & 4.18  & 0.89  & -17.07  & 6.93  & 3.39  & -18.35  & 7.24  & 3.89  \\
\cmidrule{2-15}          & 300   & EM    & -1.23  & 8.90  & 0.81  & -0.67  & 4.87  & 0.24  & -0.95  & 8.76  & 0.78  & -1.26  & 8.80  & 0.79  \\
          &       & NP    & -0.65  & 9.85  & 0.97  & -0.14  & 4.91  & 0.24  & -1.91  & 8.38  & 0.74  & -0.84  & 8.89  & 0.80  \\
          &       & naive & -18.00  & 4.64  & 3.45  & -8.52  & 2.36  & 0.78  & -17.35  & 4.18  & 3.19  & -18.49  & 4.38  & 3.61  \\
\cmidrule{2-15}          & 500   & EM    & -0.68  & 6.81  & 0.47  & -0.41  & 3.67  & 0.14  & -0.60  & 6.67  & 0.45  & -0.71  & 6.73  & 0.46  \\
          &       & NP    & -0.29  & 7.68  & 0.59  & -0.04  & 3.74  & 0.14  & -1.36  & 6.76  & 0.48  & -0.40  & 6.99  & 0.49  \\
          &       & naive & -17.95  & 3.93  & 3.38  & -8.48  & 1.81  & 0.75  & -17.26  & 3.48  & 3.10  & -18.45  & 3.67  & 3.54  \\
    \bottomrule
    \end{tabular}
}
    \label{tab:bi-normal-1}
	\begin{tablenotes}
    \item The abbreviations are identical to Table \ref{tab-note} except that 
    $\Delta$ROC(0.2), $\Delta$AUC, $\Delta$Youden, and $\Delta$pAUC represents the differences of the corresponding quantities for the two biomarkers.  
    All values have been multiplied by $10^2$.
	\end{tablenotes}
	\label{tab-note-2}
\end{table}

\section{Real Data Analysis}\label{chap_realdata}
In this section, we apply the proposed method to the malaria data  collected from a cross-sectional survey on parasitemia and fever of children under one year old, in a village in the Kilombero district of Tanzania \citep{kitua1996plasmodium}. 
Our primary focus is to assess the performance of parasite levels in blood, recorded in the dataset, for diagnosing malaria.

The dataset can be divided into two groups: $R=0$ and $R=1$. The individuals in the $R=0$ group were collected during the dry season (July-December), when malaria prevalence is low. All individuals in this group are non-malaria cases ($G=0$). In contrast, the $R=1$ group includes individuals collected during the wet season (January-June), when malaria prevalence is high. This group contains a mixture of malaria ($G=1$) and non-malaria ($G=0$) cases \citep{qin2005semiparametric, yu2019maximum}.
Here, we treat $R$ as the reference standard, meaning individuals from the $R=1$ group (wet season) are classified as nominal malaria cases, while individuals from the $R=0$ group (dry season) are classified as non-malaria. 

The dataset consists of $n=81$ subjects with positive parasite levels from the $R=0$ group and $m=211$ subjects with positive parasite levels from the $R=1$ group. This setup aligns with model (\ref{mix_model}) where $\pi_0 = 1$ and $\pi_1 = 0.677$. It is important to note that $\pi_1 = 0.677$ is estimated from the ratio of malaria patients to fevered patients \citep{qin2005semiparametric}.

\begin{figure}[htbp]
    \centering
\includegraphics[scale=0.48]{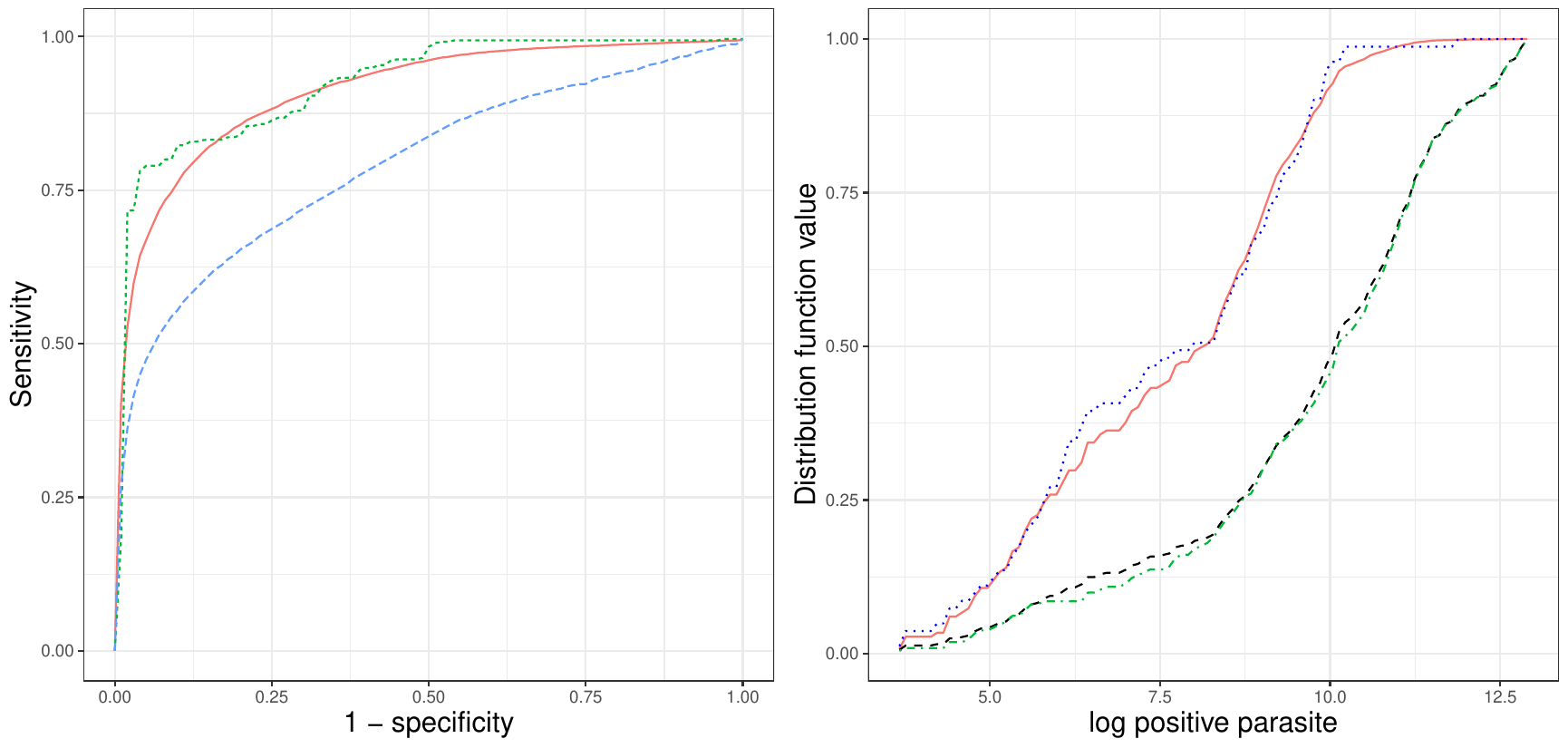}
   \caption{Left panel: The estimated ROC curves for the proposed method (red solid), the fully nonparametric method (green dotted), and the naive method (blue dashed).
Right panel: The estimated cumulative distribution functions for $F_0$ (red solid) and $(1-\pi_1)F_0 + \pi_1 F_1$ (black dashed), along with the empirical distribution functions of the nominal control group ($R=0$, blue dotted) and the nominal case group ($R=1$, green dash-dotted).}
\label{plot:ROC}
\end{figure}

We perform the ROC curve analysis using the proposed method, with the logarithm of the parasite level  as the biomarker of interest.
When implementing Algorithm \ref{algo-EM-1}, the tuning parameter $\nu$ is selected through five-fold cross-validation. The estimated cumulative distribution functions are shown in the right panel of Figure \ref{plot:ROC}, and they exhibit similar patterns to the empirical distribution functions for the two groups, $R=0$ and $R=1$. The left panel of Figure \ref{plot:ROC} presents the estimated ROC curves from the proposed method, the fully nonparametric method \citep{sun2024nonparametric}, and the naive method. While the trends are similar, the ROC curve estimated by our method is smoother compared to the one from the fully nonparametric method. This result aligns with our expectation, as the proposed method incorporates a roughness penalty for the density ratio.

\begin{table}[htbp]
\centering
\caption{The estimates of the parameters of interest using three methods}
\setlength\tabcolsep{16pt}
\begin{threeparttable} 
\small
 \begin{tabular}{ccccc}
    \toprule
     method  & ROC(0.2) & AUC   & Youden & pAUC \\
    \midrule
    proposed    & 0.856  & 0.911  & 0.672  & 0.849  \\
    fully nonparametric    & 0.839  & 0.932  & 0.739  & 0.847  \\
    naive & 0.653  & 0.793  & 0.463  & 0.647  \\
    \bottomrule
	\end{tabular}
	\begin{tablenotes}
	\centering
	\item The names of the parameters are identical to those in Table \ref{tab-note}.
	\end{tablenotes}
\end{threeparttable}
\label{tab:app-res1}
\end{table}

The estimates of parameters of interest for the three methods are shown in Table \ref{tab:app-res1}. Our method and the fully nonparametric method yield similar results for all parameters of interest. However, the Youden’s index estimate using our method is smaller than that from the fully nonparametric method. As indicated in Table \ref{tab:uni-normal-1} from the simulation, the fully nonparametric method tends to overestimate  Youden’s index. Table \ref{tab:uni-normal-1} also shows that the estimates from the naive method are significantly smaller than those obtained from the other two methods. This underestimation in the naive method is due to its failure to account for label contamination, an issue previously identified by \citet{sun2024nonparametric}.



\section*{Supplementary material}
\label{SM}
The supplementary material includes additional results from numerical studies, regularity conditions for asymptotic properties, further discussions on the proposed EM algorithm, and proofs of the propositions and theorems presented in the main paper.

\bibliographystyle{apalike}
\bibliography{ref1.bib}

\end{document}